\documentclass[aps,twocolumn]{revtex4}
 \usepackage[utf8x]{inputenc}
\usepackage{fontenc} 
\input{epsf}
\usepackage{epsfig}
\usepackage[]{graphicx} 
\usepackage{dcolumn} 
\usepackage{bm} 
\usepackage{slashed}
\usepackage{float}
\usepackage{amsmath,amssymb,amsfonts}
\usepackage{hyperref}
\usepackage{grffile}
\usepackage{fancybox}
\usepackage{setspace}
\usepackage{wrapfig}
\usepackage{lipsum}
\usepackage{color}

\usepackage[usenames,dvipsnames]{xcolor}

\newcommand{\up}[1]{u_\textbf{p}^s}
\newcommand{\vp}[1]{v_\textbf{p}^s} 
\graphicspath{ {figures/} }

\begin{document}
 
\title{Possible Universal Limit for Valence Parton Distributions}

\author{Christopher Leon, Misak M. Sargsian }

\affiliation{
Department of Physics, Florida International University, Miami, FL 33199 USA}

\date{\today}

\begin{abstract} 
We report the observation of the existence of a possible universal limit for valence parton distributions that should exist 
once partonic degrees of freedom are relevant for high energy scattering from
strongly interacting bound systems
like a nucleon, meson or a few nucleon system at very short distances.
Our observation is based on the notion that the Bjorken x weighted valence parton distribution function has a peak, $x_p$,
that characterizes the average momentum fraction carried out by the valence quarks in the system.
Within the
residual mean-field model of the valence quark distribution we found that $x_p$ has an upper limit:
$x_p \leq  {1\over 2(n_V-1)}$, where $n_V$ is the number of valence quarks which can be considered in 
the cluster embedded in the strongly interacting environment of the bound system.
The existence of such a limit imposes a new constraint on choosing the starting resolution scale $Q_0$ for PDFs.
Our prediction   for the nucleon is that  $x_p\mid_{Q\to Q_0}\le{1\over 4}$, which is 
in agreement with all the available  valence PDFs  that employ the standard approach for selecting starting  $Q_0$.
We also demonstrate  how the existence of this limit can be used to check the onset of quark-clusters
in short range nucleon correlations in nuclei. 
\end{abstract}
\maketitle

\section{Introduction}

Valence quarks are one of the most important constituents of hadrons, defining 
their baryonic number  and  representing  ``effective" fermions interacting mutually and 
with the hadronic interior.  These interactions being non-perturbative pose significant challenges in their 
description within Quantum Chromodynamics (QCD). One of the quantities representing a testing ground for a non-perturbative QCD description of 
valence quarks are the partonic distribution functions (PDFs). During the last several decades there have been
extensive efforts in the phenomenological 
extraction of valence PDFs from the analysis of deep inelastic scattering (DIS) from hadronic targets
and Drell-Yan processes\cite{cteq18,cj15,cteq14}.  There are few  empirical constraints on the behavior 
of valence PDFs and possible new constraints  are important for the precision of the  extraction of PDFs in global analyses.  
Theoretical modeling  of valence PDFs and comparing them with phenomenological PDFs is one direction that 
allows the
introduction of additional constraints,  progressing the understanding  of the valence  structure of hadrons.

\begin{figure}[ht]
\vspace{0.1cm}
\includegraphics[width = 0.43 \textwidth]{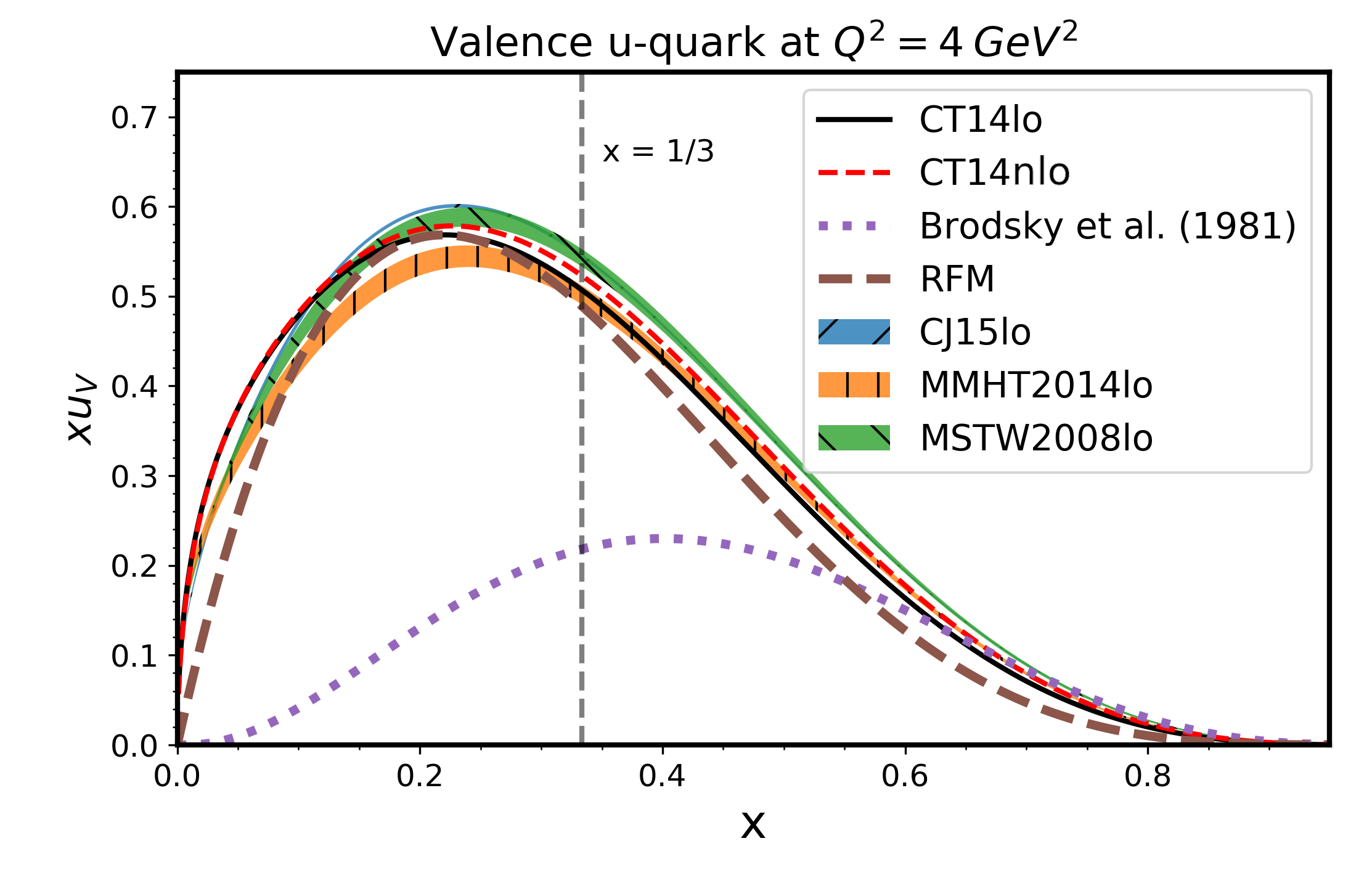}
\vspace{-0.6cm}
\caption{Peak position of $x u_V(x)$ distribution compared with prediction of different models calculating 
valence  PDFs.  }
\vspace{-0.2cm}
\label{upeak}
\end{figure}

Even though the shapes of the valence PDFs are not observables, many theoretical models however predict 
specific shapes at different ranges  of Bjorken x.  The feedback of such modeling is that their predictions  are used 
in  constructing the ansatzes for PDF parameterizations. For example, a $(1-x)^{N}$ asymptote at large x is predicted 
from the quark counting rule, according to which   $N = 2n - 3 + 2|\lambda_h-\lambda_i|$, where $n$ is the number of valence quarks and 
$\lambda_h$ and $\lambda_i$  are the helicities of the hadron and struck quark, respectively.  Similarly, the $x^s$ behavior for valence quarks is expected at small x
following from dominance of Regge dynamics with exchanged spin $s=0.5$. Both of these predictions are used in the ansatzes of phenomenological PDFs. 
\begin{figure}[ht]
\includegraphics[scale=0.15]{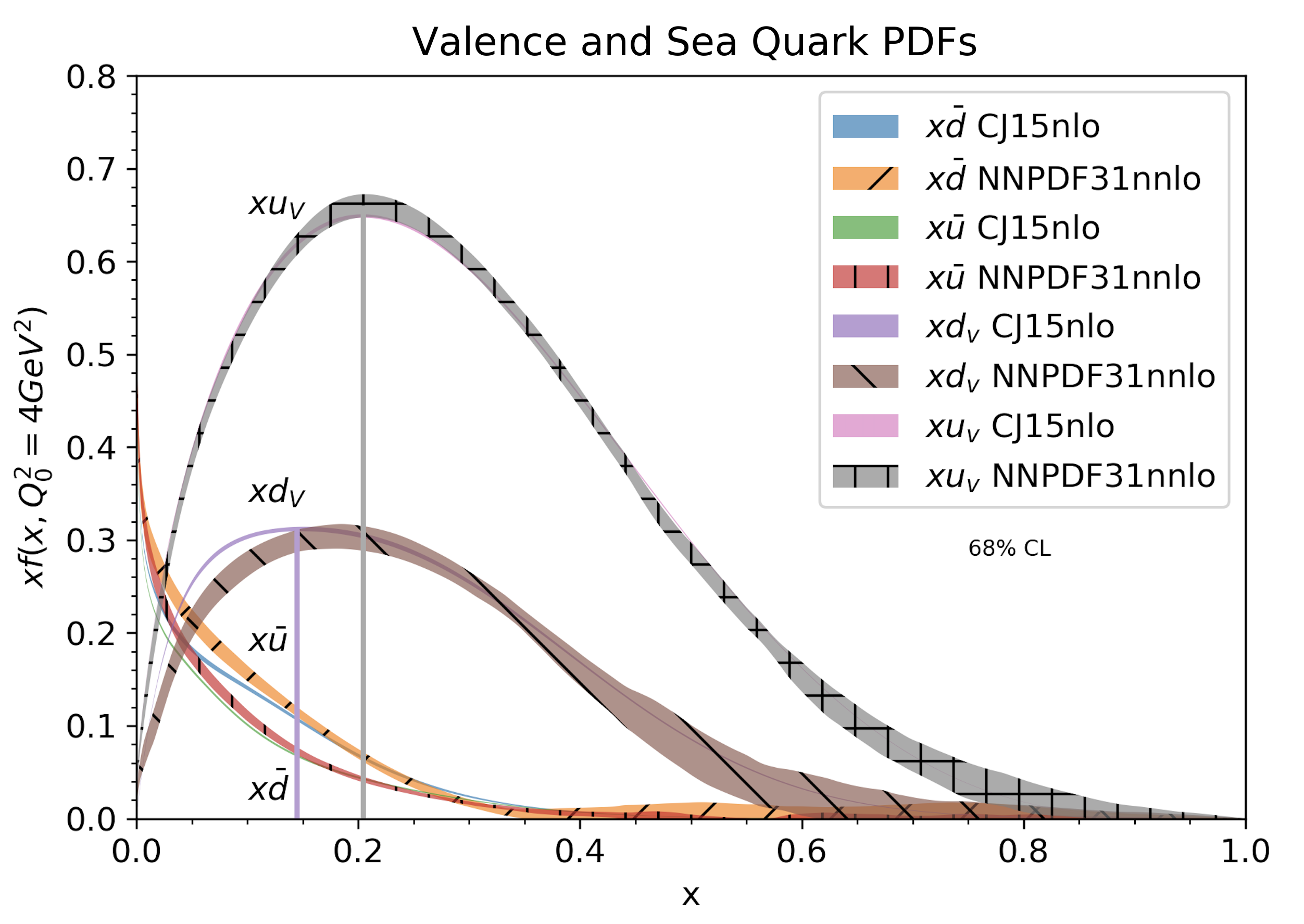}
\vspace{-0.8cm}
\caption{The valence and sea PDFs for nucleon using the CJ15nlo\cite{cj15} and NNPDF31nnlo\cite{NNPDF:2017mvq} sets evaluated at $Q_0^2 = 4 \: GeV^2$. Error bands show the Hessian 68 \% confidence level.  }
\label{peaksPDF}
\end{figure}
In the present work we focus on another unique characteristic of valence PDFs, that is the x-weighted distribution has a distinguishable 
peaking property for all approximation orders, factorization, and renormalization scales (Fig.\ref{peaksPDF}).  The QCD evolution 
shifts the peak  position of the distribution at smaller x values resulting in an interesting correlation between $x_p$ and the height of the peak in the form\cite{Leon:2020nfb}:
\begin{equation}
x_p(Q^2)q_V(x_p,Q^2) = h(x_p)  = Ce^{D x_p(Q^2)},
\vspace{-0.2cm}
\end{equation}
where $C$ and $D$ are constants.
At $Q^2$ large enough that quark degrees of freedom are relevant, and  
small enough  such that  the QCD evolution does not shift the peak position substantially towards smaller x, the peak position 
within the partonic picture characterizes the average momentum fraction carried by the interacting parton. 
For the nucleon, the naive expectation (e.g. Ref.\cite{Halzen:1984mc}) is that $x_p={1\over 3}$, indicating that each parton 
carries one third of the overall longitudinal momentum of the nucleon. Such a  result is also expected for a simple non-relativistic constituent quark model.
However, a relativistic description of a three-quark system on the Light-Front (LF) deals with more complicated dynamics of momentum sharing between quarks, resulting in,  for example \cite{Brodsky:1981jv}, a position of the peak at $x_p = 0.2$ (see Fig.\ref{upeak}). 

Moreover, as Fig.\ref{peaksPDF} shows, the position of the peak for valence $d$ quarks is systematically lower than that of the $u$-quarks. 
To the best of our knowledge this inequality first was addressed within the di-quark model\cite{Close:1988br,Roberts:2013mja,Lu:2022cjx}  in which  it was observed that  the  magnitude of $x_p$ is related to the mass of the recoil di-quark, $m_d$, in the form: $x_p \approx (1- {m_d\over m_N})$. In this picture 
the empirical observation that $x_p^d < x_p^u$ follows from the expectation that  the vector di-quark  being a spectator   in the valence 
d-quark distribution is  heavier than the scalar di-quark for the case of valence u-quark distribution.  In the di-quark model, however, one needs 
a dynamical model generating masses for di-quarks $\sim 500-700$~MeV/c.  In our spectral function  approach described in the text, we give a different explanation of this difference based on the difference in masses of the residual systems of hadrons.

This article is arranged as follows: in Sec. \ref{sect-spec}
we present a spectral function approach to modeling hadrons, outline the assumptions and derive an expression for the valence PDF in the residual mean-field model. In Sec. \ref{sect-analytic}, we present predictions, specifically for $x_p$ and bound it with an upper limit for all parameterizations. This limit is tested using existing phenomenologically extracted PDFs. Finally, Sec \ref{sect-summary} contains a summary and outlook. The appendices contain technical calculations and a numerical investigation of $x_p$ with different parameterizations.

\section{Spectral Function Approach in Description of Valence Quark Distributions}\label{sect-spec}

Recently we developed a new approach\cite{Leon:2020bvt} in  which the dynamics of the valence quarks in the nucleon is described through a spectral function framework that distinguishes between the
massless valence quark cluster and a residual nucleon system, the latter being characterized by the residual mass distribution $m_R$.
The model resulted in a prediction that the above discussed peak position of the valence PDFs are defined according to:
\begin{equation}
x_p = {1\over 4} (1 - {m_R\over M_N}),
\label{xpresmod}
\vspace{-0.0cm}
\end{equation}
where $M_N$ is the mass of the nucleon. Due to the expected different masses for the residual system, depending on whether the u- or d- valence quark is being 
considered, the model predicted $x_p^d < x_p^u$ in accordance with  empirical data (Fig.\ref{peaksPDF}). Another interesting 
consequence  of relation (\ref{xpresmod}) is that it predicts an upper limit  of $x^{max}_p = {1\over 4}$.  

To generalize the above relation for any hadron (including also nuclei in the extreme condition of quark-cluster correlations), 
we present the derivation of partonic distribution function in leading-order approximation for a system consisting of n- massless valence quark cluster 
embedded in a strongly interacting bound system. 

\subsection{The valence PDF of a ``hadron" containing of  n-valence quarks}

\begin{figure}[h]	
\includegraphics[width = 0.4 \textwidth]{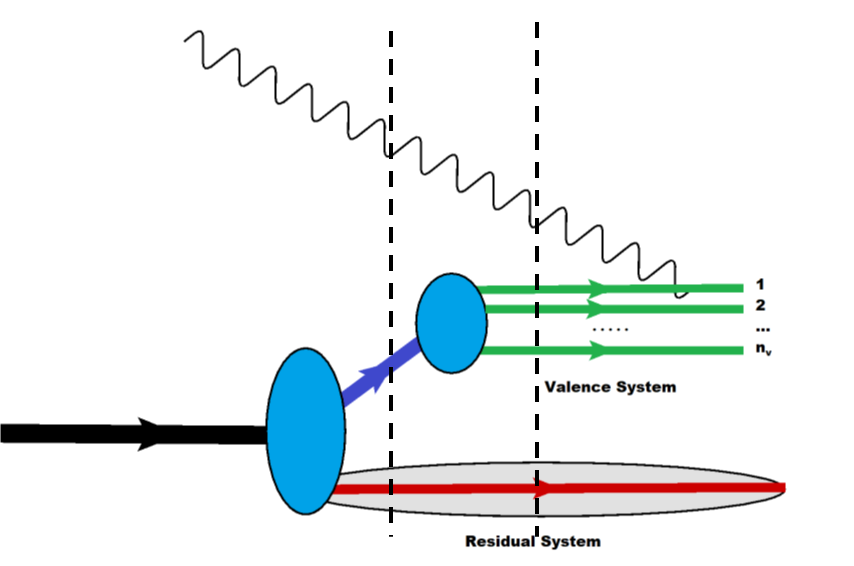}
\caption{Light cone time ordered diagram of DIS. Intermediate states are identified by the vertical dashed lines. }
\label{DIS-figure}
\end{figure}

In the spectral function approach we separate a ``hadron" containing of 
n-valence quarks  into a  cluster of n-valence quarks and a residual system consisting of  sea-quarks and gluons  (see Fig. \ref{DIS-figure}). In this scenario 
the valence quarks are embedded in the residual field, thus we refer it as residual mean-field (RMF) model, which was first applied to the case of the nucleon \cite{Leon:2020cev, Leon:2020bvt}.  In the current generalization to the n-valence quarks the final result, in addition to the nucleon, can be applied to mesons as well 
as a six-quark system that can be formed due to the strong overlap of two nucleons in the deuteron. Furthermore, by hadron we will also mean strongly overlapped two-nucleon system.

In the current approach rather than taking a full or truncated Fock  expansion of the hadron, $h$, we model its ket state via:
\begin{align}\label{hadron-ket}
|h\rangle = \psi_{n_Vq} |n_Vq\rangle \otimes \psi_{VR} |VR\rangle , 
\end{align}
where $\psi_{n_Vq}$ and $ \psi_{VR}$ are the light-front wave functions (LF) of the n-valence  cluster  ($n_Vq$) and  the
cluster-residual $VR$ system, respectively. 
 
Within our framework the LF wave functions can be related to valence PDFs if one calculates the $F_2(x,Q^2)$ structure function that enters in the
cross section of deep inelastic scattering (DIS), $eh \rightarrow e' X$. 
We consider this reaction in the Drell-Yan-West reference frame 
\begin{eqnarray}
p^\mu &= & \left(p^+, p^- ,\mathbf{p}^\perp\right) =  \left(p^+, \frac{m_H^2}{p^+}, \mathbf{0}_\perp
\right), \nonumber \\  
q^\mu &= & \left(q^+, q^- ,\mathbf{q}^\perp\right) = \left(0, \frac{2p \cdot q}{p^+}, \mathbf{q}_\perp\right),
\label{reframe}
\end{eqnarray}
where $p^\mu$ and $q^\mu$ are four-momenta of the incoming hadron and virtual photon in light cone (LC) co-ordinates, with 
$Q^2 = - q^2 =  |\mathbf{q}_\perp|^2$,  $x = \frac{Q^2}{2p\cdot q}$ and $m_H$ being the mass of the hadron.

The hadronic tensor is given by:
\begin{align}
&W^{\mu\nu} (x, Q^2)   =  \frac{1}{4\pi m_H} \sum_{ q, h_i} 
\int  \delta(k_R^2- m_R^2)   \frac{d^4 k_R}{(2\pi)^3} \nonumber \\
& \times \delta(k^{\prime,2}_1- m_1^2)   \frac{d^4 k^\prime_1}{(2\pi)^3} \prod_{i=2}^{n_V} \delta(k_i^2- m_i^2)   \frac{d^4 k_i}{(2\pi)^3} \nonumber \\
& \times (2\pi)^4  \delta^{(4)}(P+q - k^\prime_1 - \sum_{i=2}^{n_V} k_i -k_R)  A^{\mu \dagger} A^\nu,
\label{Wmunumodel}
 \end{align}
where the amplitude, $A^\mu$, is that of the scattering process depicted in Fig. \ref{DIS-figure}.  Here, $k_i^\mu$, $m_i$ and $h_i$ are 4-momentum, mass, and helicity respectively, with the $i$ index denoting the valence quarks from $i=1, 2, \ldots, n_V$, with $i=1$ being the struck quark before interacting with the photon and $i=1'$ denoting it after. The index $i=R, V$ label the residual and valence subsystem, respectively. 

The Lorentz invariant phase space for a single particle in light front co-ordinates is:
\begin{align}\label{LIPS}
\delta(k^2-m^2) d^4 k = 
\left.\frac{dx d^2 \mathbf{k}_\perp}{2x} 
\right \vert_{k^- = 
\frac{k_\perp^2 +m^2}{k^+}}.
\end{align}
In the Bjorken limit, $Q^2 \rightarrow \infty$ with $x = \frac{Q^2}{2 p\cdot q}$ fixed:
\begin{align}\label{dirac-4-conservation}
&\delta^{(4)}(P+q - k^\prime_1 - \sum_{i=2}^{n_V} k_i -k_R) \nonumber\\ &=\frac{x_1}{ p_N \cdot q}  \delta(x_1 - x)  \delta \left(1 -x_1 -\sum_{i=2}^{n_V} x_i -x_R \right) \nonumber \\
&\times  \delta^{(2)}\left(\sum_{i=1}^{n_V} \mathbf{k}_{i, \perp} + \mathbf{k}_{R, \perp}\right).   
\end{align}

.

For the calculation of the amplitude $A^\mu$ in Ref\cite{Leon:2020bvt} we developed effective light-front diagrammatic 
rules in which we introduced LFWFs through the phenomenological vertices and light-front energy denominators in the same form which enters
in LF equation of bound states. For the n-valence cluster ($n_V q$) and  cluster-residual system (VR) LFWFs are defined as:
 \begin{eqnarray}
&& \hspace{-0.4cm}  \psi_{VR} (x_V, \mathbf{k}_{R, \perp}, x_R, \mathbf{k}_{V, \perp})   =   \frac{\bar{\chi_V  }\bar{\chi_R} \Gamma^{h \rightarrow VR} \chi_h}
{m_h^2 - \frac{k_{V, \perp}^2 + m_V^2}{x_V} - \frac{k_{R,\perp}^2 + m_R^2}{x_R}} \nonumber \\
&&  \hspace{-0.4cm} \psi_{n_Vq} (\{\beta_i, \mathbf{k}_{i, \perp}, h_i \}_{i=1}^{n_V})   =      \frac{\prod\limits_{i=1}^{n_V} \bar{u}(k_i,h_i)  \Gamma^{V \rightarrow n_Vq} \chi_V  }
{m_V^2 - \sum\limits_{i=1}^{n_V} \frac{ k_{ i, \perp}^2 +m_i^2 }{\beta_i} },  
\label{LFWFs}
\end{eqnarray}
where $(x_V,k_{V,\perp})$ and $(x_R,k_{R,\perp})$ characterize LC momenta of the cluster and the residual system and 
$\{\beta_i, \mathbf{k}_{i, \perp}, h_i \}_{i=1}^{nV}$  denotes the LC momenta and helicities of the n-valence quarks in the cluster.
In the above defined LFWFs  $\Gamma^{h \rightarrow VR}$ and  $\Gamma^{V \rightarrow n_Vq}$   are phenomenological vertices
describing the transition of the hadron to the cluster-residual system and 
the valence cluster to n-valence quarks. These vertices absorb all 
the complexities of LFWS (for example null-modes) that  in principle could be calculated in an almost infinite number coupled channel 
equation for LF Fock state wave functions (see e.g. Ref.\cite{Brodsky:1997de}).  In our approach these wave functions are found 
phenomenologically by parameterizing them and then verifying their parameters in different QCD processes in which the same  LFWFs enter.

In leading order we can relate the amplitude to the LFWFs as follows:
\begin{align}
A^\mu &= \sum_{h_1, h_V}\bar{u}(k_1',h_1')   (ie_1 \gamma^\mu)   u(k_1,h_1)  
\frac{\psi_{VR}
}{x_V} \frac{\psi_{n_V q}
}{\beta_1}.
\label{Amu3}
\end{align}
Here $e_1$ is the charge of quark $i=1$.The LFWFs have dependencies of:
\begin{align}
\psi_{VR} = \psi_{VR}(x_V, \mathbf{k}_{R, \perp}, x_R, \mathbf{k}_{V, \perp})    
\end{align}
\begin{align}
\psi_{n_V q}=\psi_{n_V q}
(\{\beta_i, \mathbf{\tilde{k}}_{i, \perp}, h_i \}_{i=1}^{n_V })    
\end{align}
where 
\begin{align}
 \mathbf{\tilde k}_{i, \perp} &= \mathbf{k}_{i, \perp} - \frac{x_i}{x_V} \mathbf{k}_{V, \perp}, \\
 \beta_i &= \frac{x_i}{x_V}  \  \ \mbox{(i = $1$,...$n_V$.)}.
 \label{rel_beta}
  \end{align}
Eq. \ref{Amu3} simplifies for $\mu = +$ , resulting in
\begin{align}\label{plus-amplitude}
A^+ &= 2ie_1 \sum_{h_1, h_V} \psi_{VR} \psi_{n_V q}.
\end{align}
In the considered reference frame we can relate  the second structure function to the hadronic tensor:
\begin{align}
F_{2}(x,Q^2)  & =  {m_H (p\cdot q)\over (p^+)^2} W^{++} = {m_H Q^2\over 2x (p^+)^2}W^{++}.
\label{F2_A}
\end{align}
The non-negativity condition on the `+' momentum in Eq.(\ref{reframe})  eliminates the vacuum contribution where the photon fluctuates 
into a quark/anti-quark pair, $\gamma^* \rightarrow q \bar{q}$.
Substituting  Eq.'s  (\ref{LIPS}),  (\ref{dirac-4-conservation}) and  (\ref{plus-amplitude}) into $W^{++}$ in Eq. (\ref{Wmunumodel}), from Eq. (\ref{F2_A}) one obtains:
\begin{align}
&F_2(x) = \sum_q \sum_{h_i} \int^{Q^2} [dx]_{n_V+R} [d^2\mathbf{k}_\perp]_{n_V+R}  \nonumber \\
&\times e_q^2 x_1 \delta(x_1 - x) |\psi_{3q}
|^2 |\psi_V 
|^2,
\end{align}    
where
\begin{align}
[dx]_{n_V+R} &= \prod_{i=1,...n_V, R} \frac{dx_i}{x_i} \delta \left(1-\sum_{i=1,...n_V, R}x_i\right) 
\end{align} 
\begin{align}
&[d^2 \mathbf{k}_\perp]_{n_V+R} \nonumber\\
&= \prod_{i=1,...n_V, R} \frac{d^2 \mathbf{k}_{i, \perp}}{16 \pi^3} 16 \pi^3\delta^{(2)}\left(\sum_{i=1,...n_V, R}\mathbf{k}_{i, \perp}\right) . 
\end{align}
Using the leading order relation of $F_2(x,Q^) =  \sum_i e_i^2 xf_i(x,Q^2)$, one can then relate the valence PDF through the LFWFs as follows:
\begin{align}
&f_q(x)= \sum_{h_i} \int^{Q^2} [dx]_{n_V+R} [d^2\mathbf{k}_\perp]_{n_V+R}    \nonumber \\
& \times \delta(x_1 - x)|\psi_{n_Vq}
|^2 |\psi_V 
|^2.
\label{fq_mf}
\end{align}
Using $\mathbf{\tilde{k}}_{i, \perp}$ for the valence quarks allows us to factorize the transverse integral:

 \begin{eqnarray}
    & &\hspace{-0.4cm} f(x)    =   \int [dx]_{n_V+R} \delta(x_1-x)\nonumber\\
    &&\hspace{-0.2cm}\times  \int^{\tilde{Q}^2_{max}} [d^2 \tilde{\mathbf{k}}_\perp]_{n_V}
    \left |\psi_{n_Vq}(\{\beta_i, \mathbf{\tilde{k}}_{i, \perp}, h_i \}_{i=1}^{n_V })  \right|^2 \nonumber\\
    & &\hspace{-0.2cm}\times \int^{Q_{VR, max}^2} \frac{d^2\mathbf{k}_{R,\perp} }{16\pi^3} |\psi_{VR}(x_V, \mathbf{k}_{R, \perp}, x_R, \mathbf{k}_{V, \perp}) |^2,
\label{fx_I}
\end{eqnarray}
where:
\begin{eqnarray}
& & [d^2 \mathbf{\tilde{k}}_\perp]_{n_V}    =   16 \pi^3 \delta^{(2)}\left(\sum_{i=1}^{n_V} \mathbf{\tilde{k}}_{i,\perp}\right) \prod_{i=1}^{n_V}  \frac{d^2 \mathbf{\tilde{k}}_{i,\perp}}{16 \pi^3}\nonumber \\
& & \mathbf{\tilde k}_{i, \perp} =   \mathbf{k}_{i, \perp} - \frac{x_i}{x_V} \mathbf{k}_{V, \perp}, \ \ 
 \beta_i = \frac{x_i}{x_V}  \  \ \mbox{(i = $1$,...$n_V$)}.
\end{eqnarray}

\subsection{Valence PDF with Light Front Harmonic Oscillator LFWFs}
From the fact that the hadron is a bound, relativistic object we model $\psi_{n_Vq}$ through the mutually coupled 
relativistic  harmonic light-front wave functions\cite{Leon:2020bvt}, to account the effect of confinement. For $\psi_{VR}$, we use a Gaussian function with 
non-relativistic kinematics used to estimate the $z$ component of momentum.  
\begin{align}
\psi_{n_V q} & = 16\pi^3 m_H A_R \exp{ \left[- \frac{B_V}{8}\sum_{i=1}^{n_V} \frac{\tilde{k}^2_{i,\perp} + m_i^2}{\beta_i} \right]} \nonumber \\
&\times e^{\frac{n^2 m^2 B_V}{8}}  \sqrt{x_2....x_{n_V}} \\
\psi_{VR} &=  \sqrt{16\pi^3 m_H}  A_Vs \nonumber\\
&\times \exp{ \left[-\frac{B_R}{2}\left( \left(m_H x_R - m_R\right)^2  k_{R, \perp}^2 \right) \right]} \sqrt{x_R}.
\end{align}
Inserting the above wave functions into Eq.(\ref{fx_I}) one obtains  :
\begin{align}\label{f_q}
    f(x) & =  (16\pi^3m_H)^5 |A_R A_V|^2 e^{\frac{n^2 m^2 B_V}{8}} \nonumber \\
    &\times \int [dx]_{n_V+R} \delta(x_1-x) x_2....x_n \nonumber\\
    &\times \int^{\tilde{Q}^2_{max}} [d^2 \tilde{\mathbf{k}}_\perp]_{n_V}
    e^{-\frac{B_V}{4}\sum_i \frac{\tilde{k}^2_{\perp,i}+m_i^2}{\beta_i}}
    \nonumber\\
    & \times \int^{Q_{VR, max}^2} \frac{d^2\mathbf{k}_{R,\perp} }{16\pi^3} |\psi_{VR}|^2.
\end{align}
Taking the  $Q^2 \rightarrow \infty$ limit (see 
Appendix \ref{transverse-integral}) results in  :
\begin{align}
 &\int [d^2\tilde{\mathbf{k}}]_{n_V} e^{-\frac{B_V}{4}\sum_i \frac{\tilde{k}^2_{\perp,i}+m_i^2}{\beta_i}} \nonumber \\ 
    &= \frac{e^{-\frac{B_V}{4}\sum_i \frac{m_i^2}{\beta_i}}}{(16\pi^2)^{n_V-1} (B_V/4)^{n_V-1} } \frac{x_1 x_2 ...x_{n_V}}{x^{n_V}_V}.\label{transv-int-valence}
\end{align}
Similarly, for the residual transverse integral:
\begin{align}
    &\int \frac{d^2\mathbf{k}_{R,\perp} }{16\pi^3}  |\psi_{VR}|^2 \\
    &= \frac{\pi  e^{-B_Rm_H^2(x_R-\mu_R)^2} x_R }{16\pi^3} \int dk^2_{R,\perp}e^{-B_Rk^2_{R,\perp}}\nonumber\\ 
    & = \frac{  e^{-B_Rm_H^2(x_R-\mu_R)^2} x_R }{16\pi^2 B_R}, \label{transv-int-residual}
\end{align}
where $\mu_R = \frac{m_R}{m_H}$. Inserting Eq. (\ref{transv-int-valence}) and (\ref{transv-int-residual})  into Eq. (\ref{f_q}) for the valence PDF one obtains:
\begin{align} \label{RMF-model-val-PDF}
    f(x) &= \mathcal{N}_{n_V}\int [dx]_{n_V+R} \nonumber  \\
    &\times \delta(x_1 - x)\frac{x_1x^2_2...x^2_{n_V}}{x^{n_V}_V}
    e^{-B_Rm_H^2(x_R-\mu_R)^2}x_R,
\end{align}
where 
\begin{align}
\mathcal{N}_{n_V} &= \frac{(16\pi^3 m_H)^5 4^{n_V-1}|A_R A_V|^2 }{(16\pi^2)^{n_V} B_V^{n_V-1} B_R} e^{\frac{B_V n_V^2 m^2}{4}}. 
\end{align}

This can be simplified (see Appendix \ref{pdf-x-integral}) to obtain:
\begin{align}
    f(x) &= \frac{ \mathcal{N}_{n_V}}{(2n_V-3)!} \nonumber \\
    &\times \int\limits_0^{1-x}   dx_R  \frac{(1-x_R-x)^{2n-3}}{(1-x_R)^n}e^{-B_Rm_H^2(x_R-{m_R\over m_H})^2}. 
    \label{fvq}
\end{align}
\section{Analytic Predictions of the Model}\label{sect-analytic}
Obtained in Eq.(\ref{fvq}) expression uses a coupled $  n_V$-dimensional harmonic oscillator on the Light-Front which does not contain  the
hard component, which can be  generated through the hard gluon exchanges between valence quarks.  
With such a ``soft" wave function one expects that the model 
has more validity  at moderate $x$ relevant to the peak position of the $x$ weighted valence PDF distribution (see Fig.\ref{peaksPDF} and Ref.\cite{Leon:2020nfb}). We expect our model to underestimate  the high $x$  part of the valence PDFs   if they are dominated by 
hard interactions between valence quarks\cite{Brodsky:1974vy}. 

Our procedure for the case of the nucleon\cite{Leon:2020nfb}  was to fit the height and the position of the x-weighted valence PDF peak and 
thereby to evaluate parameters entering the LFWFs of Eq.(\ref{LFWFs}). Once these parameters are fixed then one can use the same wave functions in the calculation of different QCD ``objects'' like form-factors, generalized partonic distributions or transversities that can be 
accessed experimentally. This allows us to verify the universalities of LFWFs for multitude of processes at moderate range of $x\sim 0.1-0.4$.

In the present work  we focus on analytic  features  of the model that does not require a fitting.  For this,  one observes that 
for moderate $x$,  using the exponent in the integrand in Eq.(\ref{fvq}) we can evaluate it at its minimum,
$x_R \sim 1-{m_R\over m_H}$, resulting in:
\begin{equation}\label{approx_f}
\hspace{-0.2cm} f(x) \approx \frac{ \mathcal{N}_{n_V}}{(2n_V-3)! (1-\mu_R)^{n_V}} (1-x-\mu_R)^{2n_V-3}.    
\end{equation}
The derivation of the approximation in Eq. (\ref{approx_f}) utilizes the saddle point approximation, which has a correction of $\mathcal{O}(\frac{1}{B_R m_H^2}) $. Values of $B_R m_H^2$ extracted from fits to phenomenologically derived proton valence PDFs are typically $\sim 10 -50$ \cite{Leon:2020bvt}.
Using Eq. (\ref{approx_f}) one can calculate the position of the peak of $x f(x)$  to obtain:
\begin{equation}\label{approx_x_peak}
x_p = \frac{1}{2(n_V-1)} \left( 1- {m_R\over m_H} \right).
\vspace{-0.2cm}
\end{equation}
By the evolution of the DGLAP equations, there is a dependence on $x_p$ on $Q^2$. While there is no exact prescription as to what minimal $Q_0^2$ to use, one expects it to be large enough for factorization and pQCD to be valid and small enough so that the exact characteristics of the residual structure are not important. However, once fixed at a particular $Q^2$ their renormalization flow is uniquely determined.

Since the mass of the residual  system, $m_R$, is positive the above equation results in a universal upper limit of 
the peak position of valence quark PDFs  of the form:
\begin{equation}
x_p \leq \frac{1}{2(n_V-1)}.   
\label{xpeak-inequality}
\vspace{-0.2cm}
\end{equation}
This upper bound is valid for all $Q^2$. A numerical investigation into the validity of Eq. (\ref{approx_x_peak}) and (\ref{xpeak-inequality}) can be found in Appendix \ref{appendix-numerical}, where both are found to generally hold for the parameters explored. 

\begin{figure}
\includegraphics[width = 0.96\columnwidth]{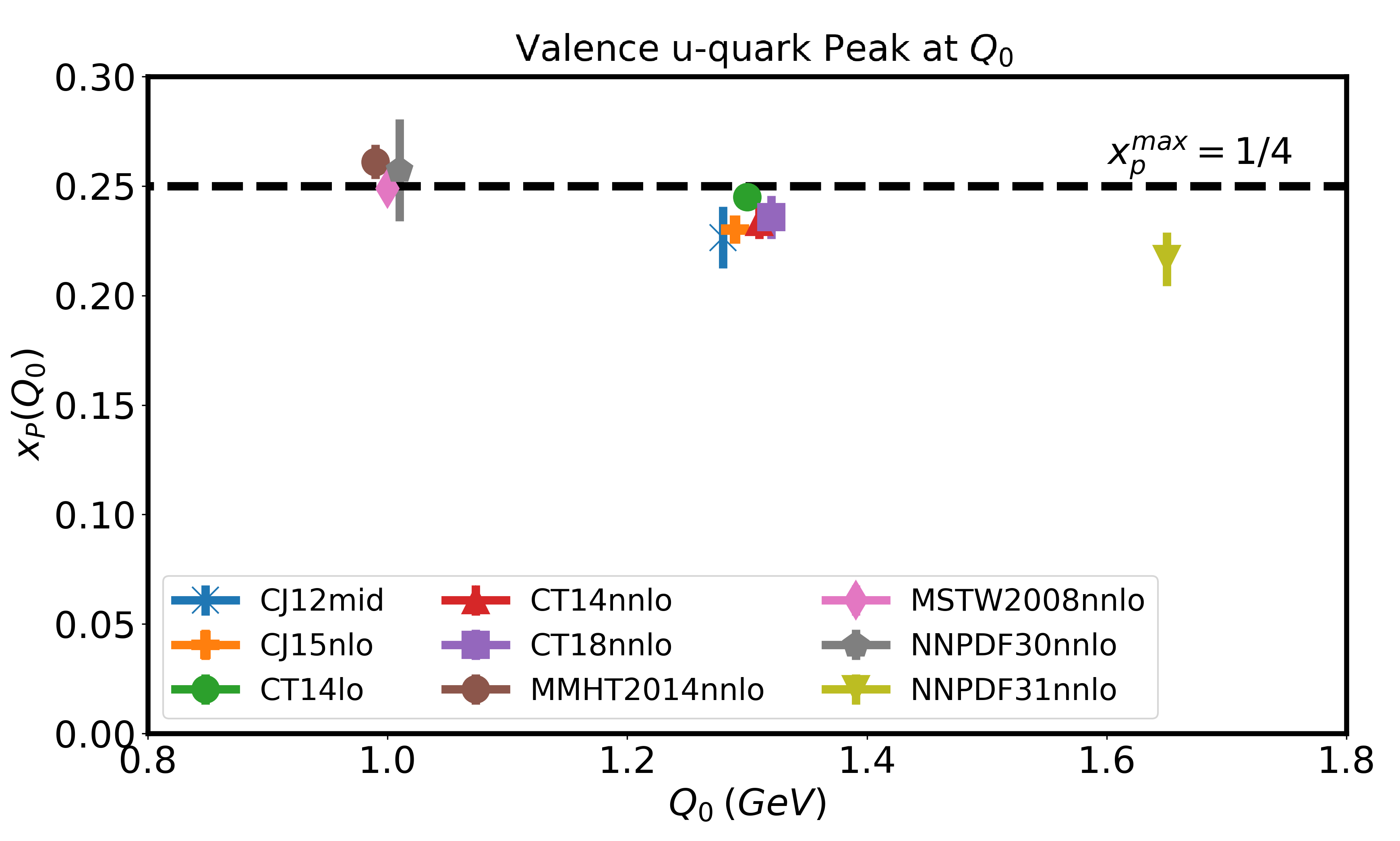}
\centering
\vspace{-0.4cm}
\caption{The up valence peak for various PDFs sets at their starting $Q_0$ \cite{cteq18, cj15, cteq14, NNPDF:2017mvq,Owens:2012bv, Harland-Lang:2014zoa,Martin:2009iq, NNPDF:2014otw}.} 
\vspace{-0.22cm}
\label{xpeaks-PDFs}
\end{figure}

For the nucleon this results in Eq.(\ref{xpresmod}) and an upper limit of $x_p^{max} = {1\over 4}$. This allows us to state that if the partonic degrees are resolved in the nucleon then the position of the peak of  the x weighted valence PDFs should not exceed ${1\over 4}$.
To check this observation, in Fig.\ref{xpeaks-PDFs} we compare the results from several valence PDFs at starting $Q_0$. Note that only those 
PDFs are chosen that apply the standard approach in defining starting $Q_0$, that is, it assumes the onset of partonic degrees of freedom 
in the  nucleon.  The figure shows 
that at the lowest possible starting $Q_0=1$~GeV (for which partonic degrees of freedom can only start to be relevant) the position of the peak 
is in agreement with the predicted upper limit of ${1\over 4}$.  For all $Q> Q_0$ the peak position is less than ${1\over 4}$, thus satisfying 
inequity (\ref{xpeak-inequality}). Note that this result is independent on the form of the ansatz in empirical PDF parameterizations. 
One may think that  $x_p^{max} = {1\over 4}$ can be a simple reflection  that, in $m_R\to 0$ limit the  model becomes like a four-massless—body problem.  
If this is the case, then   mesons will correspond to three-massless-body problem expecting  $x_p^{max}=  {1\over 3}$.  However, as 
Eq.(\ref{xpeak-inequality}) shows for mesons the model predicts $x_p < {1\over 2}$, which is in agreement with available  pion PDFs\cite{Barry:2018ort, Barry:2021osv}. Thus, the 
result is due to a nontrivial light-cone momentum sharing between valence quarks and residual system.

It is interesting that for the case of 6q-system the model predicts $x_p < {1\over 10}$, which should be compared to ${1\over 8}$ that follows from the
convolution model in which two nucleons in the deuteron did not merge into a 6q state. Thus, the shift of the peak position towards lower $x$  reaching to ${1\over 10}$ for the deuteron valence PDFs  at fixed $Q^2$ with increase of internal momentum in the deuteron will indicate the transition from NN to 6q state.

Furthermore, without fixing parameters of the LF wave functions we can calculate the analytic form of the valence PDFs at $x\to 1$.  Substituting $x\to 1-\epsilon$ 
we calculate  the integral in Eq.(\ref{fvq}) in the small $\epsilon$ limit, resulting in:
\vspace{-0.2cm}
 \begin{equation} 
f(x)  =  {\mathcal{N}_{n_V}\over (2n_V-3)!} e^{-B_Rm_H^2(1-\mu_R)^2}(1-x)^{2n_V-2}.
 \label{fvqatx1}
\end{equation} 
The above relation predicts $(1-x)^4$ behavior for nucleon valence PDFs at $x\to 1$, which should be compared with $(1-x)^3$ that follows from 
perturbative QCD arguments that require two hard gluon exchanges between three valence quarks\cite{Brodsky:1974vy}. 
It is interesting that for mesons one ovbtains  $(1-x)^2$, which is the same as the prediction from pQCD. This indicates that the observation of 
$(1-x)^2$ behavior of the pion 
valence PDF may not necessarily indicate the dominance of the hard component in the pion wave function. This observation is in agreement with that of \cite{Efremov:2009dx}. The same is true for the deuteron in which case Eq.(\ref{fvqatx1}) predicts $(1-x)^{10}$ the same behavior as predicted within pQCD\cite{Brodsky:1974vy}.

\section{Summary and Outlook}\label{sect-summary}
We introduced a residual field approach in the calculation of valence PDFs of a hadron containing $n_V$-valence quarks. In the approach 
we describe a hadron as consisting of an $n_V$-valence cluster embedded in the residual system of the hadron. Phenomenological LF wave functions are introduced, 
which do not contain a hard component and the calculation is done within the effective light-front diagrammatic method. Analyzing analytic  features of the calculated 
valence PDFs we predict that  there is a universal upper limit for the peak position of the $x$-weighted valence PDFs. Namely, for the nucleon we 
predict an upper limit for $x_p^{max} = {1\over 4}$, which is in agreement with the all available phenomenological PDFs.  This result in our view is non
trivial considering that only three-valence quarks exist in the nucleon. One obtains also $x_p^{max} = {1\over 2}$  and $x_p^{max} = {1\over 10}$  for 
meson and six-quark system respectively.

In discussing the high $x$ behavior of the PDFs one observes that soft LFWFs predict a faster drop-off for the nucleon PDFs than that of 
pQCD, while the prediction for pion and the deuteron  PDFs are the same as the pQCD prediction. 

It will be interesting if the future studies of valence PDFs, both experimentally and on the lattice, could verify  the existence of the universal 
upper limit for the nucleon $x_p^{max} = {1\over 4}$.  Such a position of the peak will be an indicator of the onset of partonic degrees of 
freedom in the nucleon. 

{\bf Acknowledgment:} This work is supported  by United States Department of Energy grant under contract DE-FG02-01ER41172


\appendix

%
\section{General Case for Transverse Integral} \label{transverse-integral}
LF relativistic harmonic oscillators were used in modelling the LFWFs. In integrating over the transverse momentum we will have integrals of the form,
\begin{align}
 I_{n_V} = \int  \prod_{j=1}^{n_V} [d^2 \mathbf{\tilde{k}}_{j, \perp}] \exp \bigg[ - \sum_{j=1}^{{n_V}} B_j \frac{\tilde{k}_{j, \perp}^2}{x_j} \bigg].  
\end{align}
We now take $Q^2 \rightarrow \infty$ in the integral bounds. The $\tilde{k}_x$ and $\tilde{k}_y$ integrals are identical, thus we can take it as the square of an ${n_V}$-dimensional integral:
\begin{align}
&I_{n_V}   = \int  \prod_{j=1}^{n_V} \frac{d^2 \tilde{k}_{j, \perp}}{16\pi^3} 16\pi^3 \delta^{(2)}\left(\sum_{j=1}^{{n_V}}\mathbf{\tilde{k}}_{j, \perp}\right)   \nonumber \\
&\times \exp \bigg[ - \sum_{j=1}^{{n_V}} B_j \frac{\tilde{k}_{j, \perp}^2}{x_j} \bigg] \nonumber \\
&=  \frac{1}{(16\pi^3)^{n_V-1}} \left( \int \prod_{j=1}^{n_V} d \tilde{k}_{j}  \delta(\sum_{j=1}^{{n_V}} \tilde{k}_{j})  \exp \bigg[  - \sum_{j=1}^{{n_V}} B_j \frac{\tilde{k}_{j}}{x_j} \bigg] \right)^2.
\end{align}
Now, we make use of the fact that we can take a Fourier decomposition of Dirac delta: $\delta(\sum_{j=1}^{N} \tilde{k}_{i})= \frac{1}{2\pi} \int_{-\infty}^{\infty} dz e^{iz(\sum_{j=1}^{N} \tilde{k}_{i})}$ in the above equation obtaining,
\begin{align*}
&I_{n_V}  = \frac{1}{(16\pi^3)^{n_V-1}} \\ 
&\times \left( \frac{1}{2\pi} \int_{-\infty}^{\infty} dz  \prod_{j=1}^{n_V} \int_{-\infty}^{\infty} d k_{j} \exp \bigg[ iz k_j \bigg] \exp \bigg[  - B_j \frac{k_{j}^2}{x_j}  \bigg] \right)^2.
\end{align*}
The integral in the product is just a Fourier transform of a Gaussian. Using the relation $\int_{-\infty}^{\infty} dx e^{-\alpha x^2} e^{i\omega x} = \sqrt{\frac{\pi}{\alpha}} e^{- \frac{\omega^2}{4\alpha}} $ in the above equation, one obtains:
\begin{align*}
I_{n_V}  &= \frac{1}{(16\pi^3)^{n_V-1}} \\
&\times \bigg( \frac{1}{2\pi} \int_{-\infty}^{\infty} dz  e^{- z^2 ( \sum_{j=1}^{n_V }\frac{x_j }{4B_j})}  \prod_{j=1}^{n_V}  \sqrt{\frac{\pi x_j}{B_j}}  \bigg)^2 \\
&= \frac{1}{(16\pi^3)^{n_V-1}}\bigg( \frac{1}{2\pi} \sqrt{\frac{\pi}{\sum_{j=1}^{n_V }\frac{x_j}{4B_j} }}  \prod_{j=1}^{n_V}  \sqrt{\frac{\pi x_j}{B_j}}  \bigg)^2. 
\end{align*}.
Thus, 
\begin{align}
I_{n_V}&=  \frac{1}{(16\pi^3)^{n_V-1}} \frac{(\pi)^{n_V-1}}{\sum_{j=1}^{n_V }\frac{x_j}{B_j} }  \prod_{j=1}^{n_V}  \frac{x_j}{B_j}.    
\end{align}
In Eq. (\ref{transv-int-valence}) the fact the $B_j$'s were the same for all $i$ and that $\sum_i \beta_i = 1$ was used. 

\section{The x Integration of PDFs}\label{pdf-x-integral}
Starting at Eq. (\ref{RMF-model-val-PDF}) we use $x_V = 1 - x_R$ and making the $[dx]_{n_V+R}$ term explicit: 
\begin{align}
    f(x) &= \mathcal{N}_{n_V} \int dx_1 \ldots dx_{n_V}dx_R\delta\Big(1-x_R-\sum^{n_V}_{i=1} x_i\Big) \nonumber \\
    & \times \delta(x_1-x) \frac{x_2...x_{n_V}}{(1-x_R)^{n_V}}e^{-B_Rm_H^2(x_R-\mu_R)^2}.
\end{align}
The Dirac delta is taken to set $x_1$ to $x$:
\begin{align}
    &f(x) = \mathcal{N}_{n_V} \int^{1-x}_0 dx_R \int^{1-x-x_R}_0 dx_2 ...  \nonumber \\
    &\times \int^{1-x-x_R-\sum_{i=2}^{n_V-1}x_i}dx_{n_V} \frac{x_2 ...x_n}{(1-x_R)^{n_V}}e^{-B_Rm_H^2(x_R-\mu_R)^2} \nonumber \\
    &\times \delta\Big(1-x_R-x - \sum^{n_V}_{i=2} x_i\Big).
\end{align}
Now, let $y_i$  be the relative momentum fraction for the $i=2, \ldots, n_V$ subsystem:
\begin{align}
    &y_i = \frac{x_i}{x_2 +x_3 + \ldots + x_{n_V}}=\frac{x_i}{1-x_R-x}
\end{align}
Doing so, the integral factorizes,
\begin{align}
    &f(x)  = \mathcal{N}_{n_V} \int^{1-x}_0 dx_R 
    \frac{e^{-B_Rm_H^2(x_R-\mu_R)^2}}{(1-x_R)^{n_V}}
    \int^{1}_0 dy_2 y_2 \ldots\nonumber \\ 
    & \int_0^{1-\sum_{i=2}^{n_V-1}y_i} dy_{n_V} y_{n_V} \frac{(1-x_R-x)^{2(n_V-1)}}{(1-x_R)^{n_V}} \nonumber \\
    &\times \delta\Big(1-x_R-x - \sum^{n_V}_{i=2} y_i (1-x- x_R)\Big) \nonumber \\
    &=  \mathcal{N}_{n_V} J_{n_V} \int^{1-x}_0 dx_R  \frac{(1-x_R-x)^{2n_V-3}}{(1-x_R)^n}e^{-B_Rm_H^2(x_R-\mu_R)^2}, 
\end{align}
where $J_{n_V}$ is a constant independent of $x_R$ and $x$. One can show $J_{n_V} = \frac{1}{(2n_V-3)!}$ (see Appendix \ref{x-spectator-integral}). Thus,
\begin{align}
    f(x) &= \frac{ \mathcal{N}_{n_V}}{(2n_V-3)!} \nonumber \\
    &\times \int^{1-x}_0 dx_R  \frac{(1-x_R-x)^{2n_V-3}}{(1-x_R)^{n_V}}e^{-B_Rm_H^2(x_R-\mu_R)^2}
\end{align}    

\section{Valence Spectator's x Integral} \label{x-spectator-integral}
In integrating over all momentum fractions of the spectator system we get the following integral as a constant:
\begin{align}\label{x-integral-expression}
J_{n_V} &= \int^{1}_0 dy_2 y_2 \ldots \int_0^{1-\sum_{i=2}^{n_V-1}y_i} dy_{n_V} y_{n_V} \delta\Big(1- \sum^{n_V}_{i=2} y_i \Big) \nonumber\\
&= \frac{1}{(2n_V-3)!}   
\end{align}
The last line of Eq. (\ref{x-integral-expression})  can be proven with mathematical induction. For the starting case of $n_V=2$,  $I_2 =1 = \frac{1}{1!}$, Now, do the induction step by assuming Eq. (\ref{x-integral-expression}) is true up to $n_V$. Then, 
\begin{align}\label{J-nv-plus-one}
J_{n_V+1} &=     \int^{1}_0 dy_2 y_2  \ldots \nonumber \\
&\times \int_0^{1-\sum_{i=2}^{n_V} y_i} dy_{n_V+1} y_{n_V+1} \delta\Big(1- \sum_{i=2}^{n_V+1} y_i \Big)
\end{align}
For $i=3, \ldots , n_V+1$ we change the variables to the relative momentum fraction in the $3, \ldots , n_V+1$ system:
\begin{equation}
z_i = \frac{y_i}{\sum_{j=3}^{n_V+1}y_j}  = \frac{y_i}{1-y_{2}}    
\end{equation}
Then, using the relation in Eq. \ref{J-nv-plus-one}
\begin{align}
J_{n_V+1} &= \int_0^1 dy_2 y_2 (1-y_2)^{2(n_V-2)} \:  J_{n_V}  \nonumber \\
&=  \frac{1}{(2n_V-3)!} \int_0^1 dy_2 y_2 (1-y_2)^{2(n_V-2)}. \label{J-nv-plus-one-integral}
\end{align} 
The integral can be taken in terms of the Beta function, $B(m,n)$, which in turn can be expressed in terms of factorials since its arguments are integers,
\begin{align}
\int_0^1 dy_2 y_2 (1-y_2)^{2(n_V-2)} &=  B  (2, 2n_V-3) \nonumber \\
&=  \frac{ 1! (2n_V-2-1)!  }{(2n_V-1)!}.\label{beta-func-used} 
\end{align}
Thus, using Eq. (\ref{beta-func-used}) in Eq. (\ref{J-nv-plus-one-integral}), one obtains:
\begin{align}
J_{n_V+1} &=  \frac{1}{(2n_V-3)!}  \frac{ 1! (2n_V-2-1)!  }{(2n_V-1)!} \nonumber \\
&= \frac{1}{(2n_V-1)!}
\end{align}
and Eq. (\ref{x-integral-expression}) holds.

\begin{figure*}
\includegraphics[ width=0.97\textwidth ]{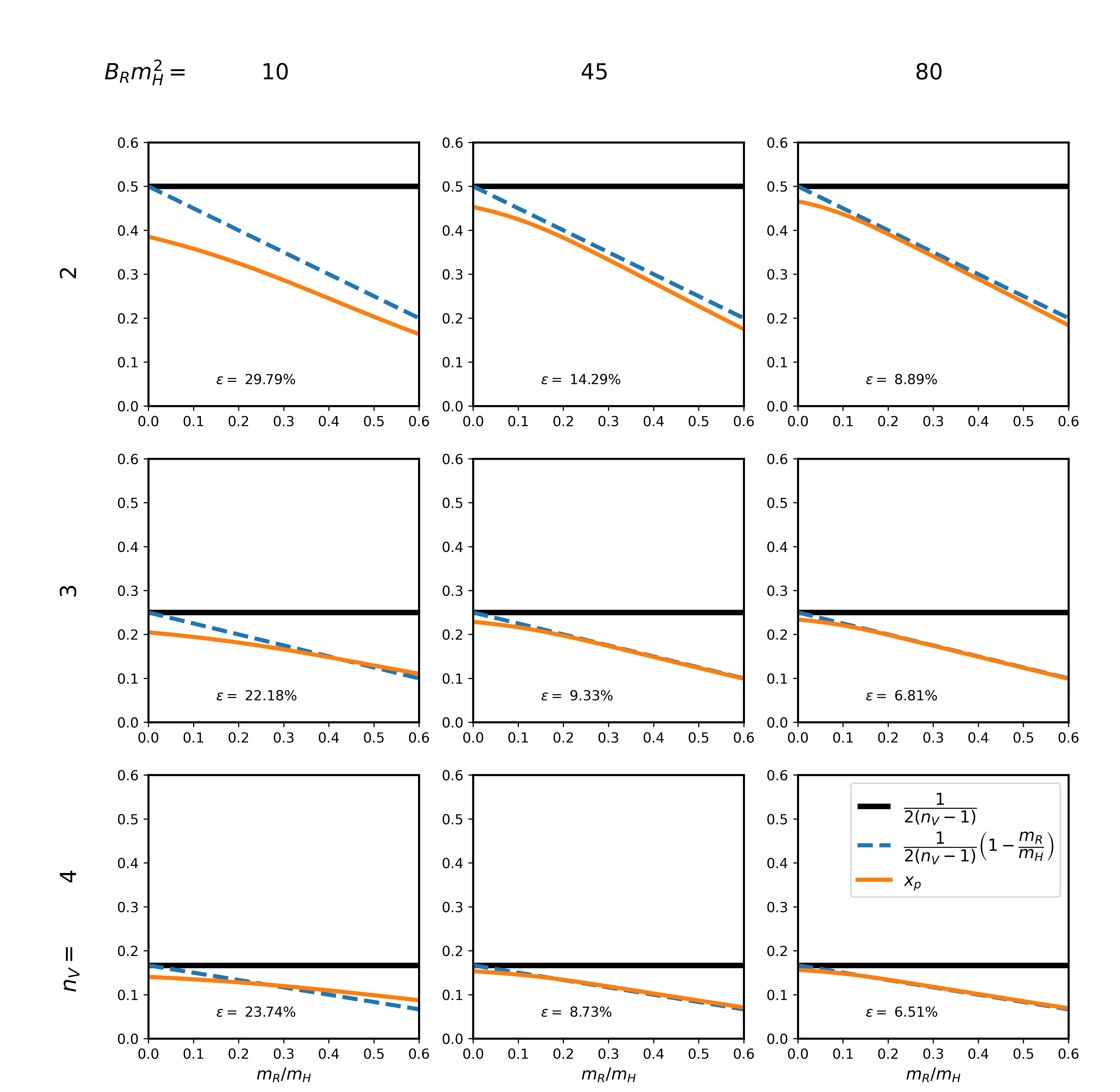}
\caption{A comparison of  $x_p$ with the approximation in Eq. \ref{approx_x_peak} for various configurations and with a mass ratio range of $0 < m_R/m_H < 0.6$. The $\epsilon $ displayed represents the largest relative error of the approximation found for the given configuration.}
\label{Fig_X_peak_approx}
\end{figure*}

\section{Numerical Investigation of x Peaking Approximation and Upper Limit}\label{appendix-numerical}

Using the saddle point approximation in  Eq. (\ref{fvq}), an approximation to the x peak, $x_p$, of the PDF was found to be:
\begin{equation}
x_p^{(sp)} = \frac{1}{2(n_V-1)} \left(1 - \frac{m_R}{m_H} \right).    
\end{equation}
From this we inferred that: 
\begin{equation}
x_p \leq \frac{1}{2(n_V-1)}.   
\vspace{-0.2cm}
\end{equation}
Here we explore the validity of the above approximation numerically by looking at the cases of  $n_V=2, 3, 4$ and $B_R m_H^2= 10, 45, 80$, the latter values being typical when the model was applied to the case of the proton \cite{Leon:2020bvt}.

Fig. \ref{Fig_X_peak_approx} shows how the peak varies with the mass ratio, $\frac{m_R}{m_H}$.  The $x_p$ value was found numerically, using the \texttt{scipy.optimize} Python library. It is interesting that for the above plots $x_p^{(sp)}$   qualitatively matches  $x_p$ well, monotonically decreasing with $m_R/m_H$ and being nearly linear. Also, in all cases investigated the upper limit $x_p ≤ \frac{1}{2(n_V-1)})$ held.


\end{document}